\documentclass[prb,showpacs,twocolumn]{revtex4}
\usepackage{amsmath}
\usepackage{graphicx}
\begin{document}

\title{Kondo Effect of Quantum Dots in the Quantum Hall Regime}

\author{Mahn-Soo Choi}%
\author{Nak Yeon Hwang}%
\author{S.-R. Eric Yang}

\affiliation{Department of Physics, Korea University, 1 5-ka
  Anam-dong, Seoul 136-701, Korea}%

\begin{abstract}
We report on the Kondo effect of quantum dots involving the precursor
of the Landau level filling factor $\nu=1$ state in the quantum Hall
regime.  We argue that pairs of degenerate single Slater determinant
states may give rise to a Kondo effect which can be mapped into an
ordinary Kondo effect in a fictitious magnetic field.  We report on
several properties of this Kondo effect using scaling and numerical
renormalization group analysis.  We suggest an experiment to
investigate this Kondo effect.
\end{abstract}
\pacs{75.20.Hr, 73.21.La, 73.23.Hk}

\maketitle

\let\up=\uparrow%
\let\down=\downarrow%
\let\Up=\Uparrow
\let\Down=\Downarrow%
\let\eps=\epsilon%
\let\veps=\varepsilon%
\newcommand\K{\,\mathrm{K}\,}%
\newcommand\T{\,\mathrm{T}\,}%
\newcommand\meV{\,\mathrm{meV}\,}%
\newcommand\ket[1]{\left|#1\right\rangle}%
\newcommand\bra[1]{\left\langle#1\right|}%

The Kondo effect, one of the most extensively studied phenomena in
condensed-matter physics~\cite{Hewson93a}, has recently enjoyed a
revival in mesoscopic systems.  Examples include quantum
dots~\cite{Goldhaber-Gordon98a,Goldhaber-Gordon98b,Cronenwett98a,vanderWiel00a,JiY00a},
quantum point contacts~\cite{Cronenwett02a}, and carbon nanotube
coupled to superconductors~\cite{Buitelaar02a}.  The main attraction
of the Kondo effect in such systems is its tunability, which makes it
possible to test various aspects of the Kondo effect that cannot be
directly investigated in bulk solids.  For example, a scattering phase
shift at the Kondo resonance in a quantum dot has been measured using
a two-path interferometer~\cite{JiY00a}.
In mesoscopic systems the role of magnetic impurities embedded in bulk
solids is played by nanoscale ``artificial atoms'', like quantum dots
or carbon nanotube (as a whole) coupled to electrodes.  The Kondo
effect arises essentially from the spin-degenerate energy levels
associated with a single unpaired electron in the artificial atom, and
is accompanied by the Kondo resonance at the Fermi level.
In quantum dots a Kondo effect may arise when an external magnetic
field induces energy degeneracy~\cite{Pustilnik00a,Eto00a,Izumida01a,Sasaki00a},
while in bulk systems a magnetic field lifts the spin-degeneracy and
leads to splitting of the Kondo resonance
peak~\cite{Meir93a,Wingreen94a}.

\begin{figure}[b]
\begin{center}
\includegraphics[width=60mm,clip=]{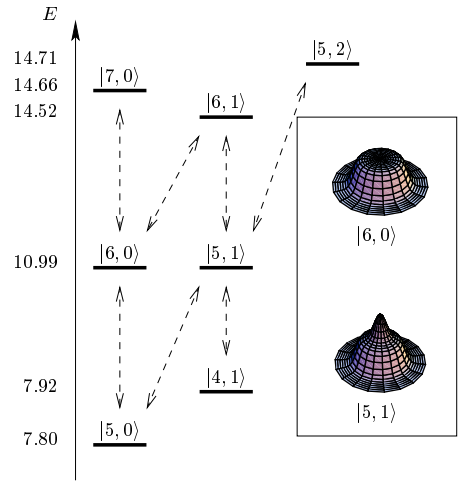}
\caption{Energy level structure of single-Slater-determinant states
  $\ket{N_\up,N_\down}$ for $N=6$.  The energies are in units of
  $e^2/\eps\ell$.  States $\ket{6,0}$ and $\ket{5,1}$ are degenerate
  ground states for certain values of the parameters (see the
  text)~\cite{YangSRE93a}.  The virtual states that are relevant to
  the Kondo effect are also Slater determinants since ground states
  are given by Slater determinants.  The arrows connect degenerate
  ground states to Slater determinant states with one more or less
  total electron number.  Inset: Density profiles of the
  maximum-density-droplet state $\ket{N,0}$ and the state
  $\ket{N-1,1}$ with one spin flipped.}
\label{Kondo:fig1}
\end{center}
\end{figure}

In this paper we report on the Kondo effect of quantum dots involving
the precursor of the Landau level filling factor $\nu=1$ state in the
quantum Hall regime.
An attractive feature of these quantum dots is that
single-Slater-determinant states are exact groundstates of the
many-body Hamiltonian under certain circumstances
~\cite{YangSRE93a,MacDonald93a,YangSRE02a}.
The Slater-determinant states, denoted by $\ket{N_\up,N_\down}$, have
$N_\up$ spin-up and $N_\down$ spin-down electrons, see
Eq.~(\ref{Kondo/SchriefferWolf:Basis}).
In this work we focus on the parameter region where $\ket{N,0}$ and
$\ket{N-1,1}$ become nearly degenerate ground states
~\cite{YangSRE93a,MacDonald93a}(see Fig.~\ref{Kondo:fig1}).  The
degeneracy of $\ket{N,0}$ and $\ket{N-1,1}$ results from the interplay
of many-body interaction, parabolic confinement energy, and (real)
Zeeman energy.
These states are stable ground states in wide regions of the parameter
space~\cite{YangSRE93a,MacDonald93a} and are easily probed in
experiments~\cite{Oosterkamp99a,Klein95a}.  We show that this pair of
degenerate many-body ground states give rise to a Kondo effect which
can be mapped into an ordinary Kondo problem with a \emph{fictitious}
magnetic field.  Also our investigation suggests that the effects of
the fictitious field may be removed by moving off the phase boundary.
Our analysis indicates that a similar effect can arise at each phase
boundary between $\ket{N_\up,N_\down}$ and $\ket{N_\up-1,N_\down+1}$
ground states.  We report on several properties of this Kondo effect
using scaling and numerical renormalization group (NRG) analysis.  We
also suggest an experiment to investigate this Kondo effect.

We first describe briefly our model for a quantum dot in a strong
magnetic field.  It consists of two dimensional electrons confined to
a finite area that is coupled to two leads.  The confining potential
is regarded parabolic,
\begin{math}
V(x,y) = \frac{1}{2} m^*\Omega^2(x^2+y^2),
\end{math}
where $ m^*$ is the effective electron mass and $\Omega$ is the
frequency scale.  A magnetic field $B$ is applied perpendicular to the
quantum dot, i.e., in the $z$-direction.  The field is assumed to be
so strong that the Landau level spacing $\hbar\omega_c$, where
\begin{math}
\omega_c=eB/m^*c
\end{math},
is sufficiently large compared with the confinement energy
$\hbar\Omega$ and the electron-electron interaction energy
$e^2/\eps\ell$ ($\ell\equiv\sqrt{\hbar c/eB}$ and $\eps$ is the
dielectric constant).  In this limit, only the lowest Landau level is
relevant.  In the symmetric gauge, the single-electron orbits of an
isolated quantum dot are labeled by the orbital angular momentum $m$
($m=0,1,2,\dots$) and spin index $\sigma=\up,\down$.  An electron in
the angular momentum state $m$ is located in a ring of the width
$\ell$ and a radius $R_m=\sqrt{2(m+1)}\ell$.  The single-particle
energy of an orbit with $m$ and $\sigma$ is given by
$\epsilon_{m,\sigma} = \gamma (1+m) + \frac{1}{2} g\mu_BB\sigma $
where $\gamma \equiv \hbar\omega_c(\Omega/\omega_c)^2$ and $g\mu_BB$
is the Zeeman splitting.

The many-body Hamiltonian of an isolated quantum dot in a strong
magnetic field is then given by
\begin{widetext}
\begin{equation}
\label{Kondo:H0}
H_0 = \sum_{m,\sigma}\veps_{m,\sigma}d_{m,\sigma}^\dag d_{m,\sigma}
+ \sum_{m_1,m_1',m_2,m_2'}\sum_{\sigma_1,\sigma_2}
U_{m_1',m_2';m_1,m_2} d_{m_2',\sigma_2}^\dag
d_{m_1',\sigma_1}^\dag d_{m_1,\sigma_1}d_{m_2,\sigma_2} \,,
\end{equation}
\end{widetext}
where $d_{m,\sigma}^\dag$ ($d_{m,\sigma}$) creates (annihilates) an
electron in the state $m,\sigma$, and $U_{m_1',m_2';m_1,m_2}$ are the
matrix elements of the electron-electron Coulomb interaction.  As
mention above, in quantum dots in a strong magnetic field the
single-Slater-determinant state of the form
\begin{equation}
\label{Kondo/SchriefferWolf:Basis}
\ket{N_\up,N_\down}
= d_{N_\down-1,\down}^\dag\cdots d_{0\down}^\dag\,
  d_{N_\up-1,\up}^\dag\cdots d_{0\up}^\dag\ket{0} \,,
\end{equation}
so-called \emph{maximum density droplet states}, can be an exact
eigenstate of the many-body Hamiltonian, Eq.~(\ref{Kondo:H0}), in a
wide range of parameters
\begin{math}
\tilde\gamma\equiv\gamma/(e^2/\eps\ell)
\end{math}
and
\begin{math}
\tilde{g}=g\mu_BB/(e^2/\eps\ell)
\end{math}
(see Figs.~1 and 2 in Ref.~\onlinecite{YangSRE93a}).

In the presence of the coupling to the leads, the many-body states in
the quantum dot is hybridized with the conduction bands of the leads.
The coupling can be considered within the tunneling model
\begin{equation}
\label{Kondo/SchriefferWolf:HT}
H_T
= \sum_{k,m,\sigma}V
\left(d_{m,\sigma}^\dag c_{k,\sigma}
  + c_{k,\sigma}^\dag d_{m,\sigma}\right) \,,
\end{equation}
where $c_{k,\sigma}^\dag$ and $c_{k,\sigma}$ are operators for
conduction electrons, and $V$ is the tunneling amplitude (we ignore
the $m$-dependence of $V$ for simplicity, see below).
We assume that there are $N$ electrons in the quantum dot in
equilibrium, ant that the Fermi level $E_F$ of the leads lies between
the successive electro-chemical potentials $\mu_{N-1}$ and $\mu_{N}$
of the quantum dot~\cite{vanHouten92a}:
\begin{math}
\mu_{N-1} < E_F < \mu_N \,.
\end{math}
Here notice that the electro-chemical potential of the quantum dot
includes the contribution from the gate voltage $V_g$ applied to the
quantum dot, which is given by
\begin{math}
\mu_N \equiv E_{N+1}^0-E_{N}^0 + eV_g
\end{math}
where $E_N^0$ is the $N$-electron ground-state energy.
Below we consider the Kondo effect in the limit
\begin{math}
\Gamma \ll E_F-\mu_{N-1}
\end{math}
and
\begin{math}
\Gamma \ll \mu_N - E_F
\end{math}
(\begin{math} \Gamma \equiv 2\pi\rho_0|V|^2.
\end{math})
We emphasize, however, that the spins involved here is not real spins
but pseudo-spins corresponding to the degenerate ground states
$\ket{N,0}$ and $\ket{N-1,1}$.

\begin{figure}[b]
\centering%
\includegraphics[width=30mm,clip=]{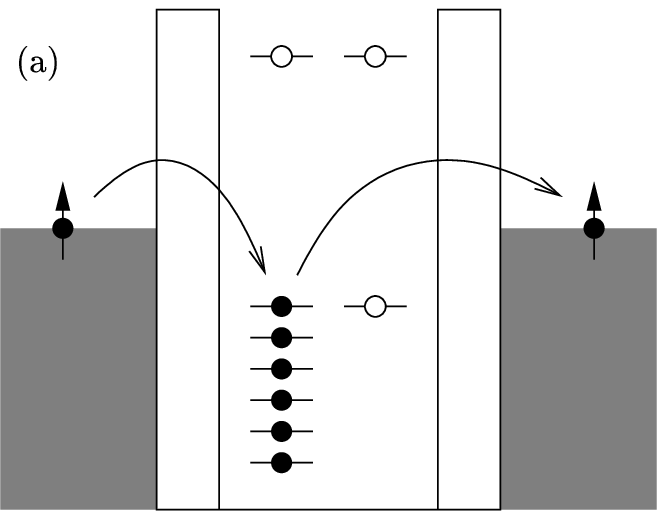}\ %
\includegraphics[width=30mm,clip=]{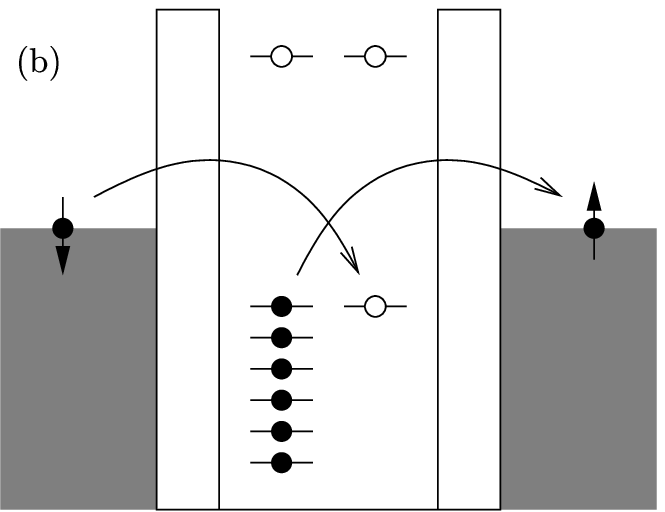}\\%
\includegraphics[width=30mm,clip=]{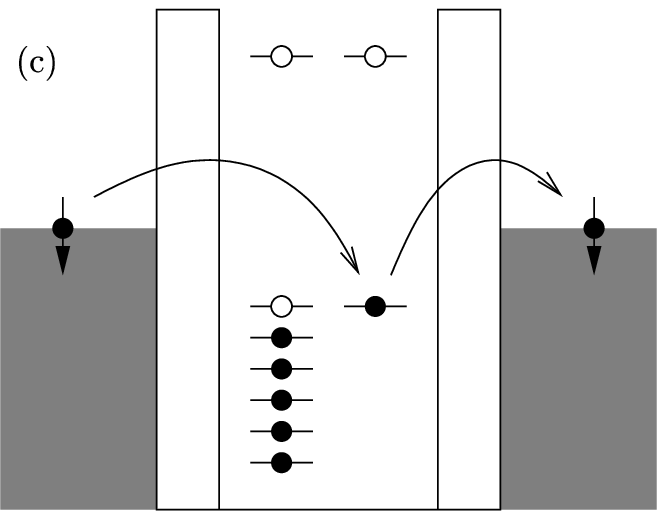}\ %
\includegraphics[width=30mm,clip=]{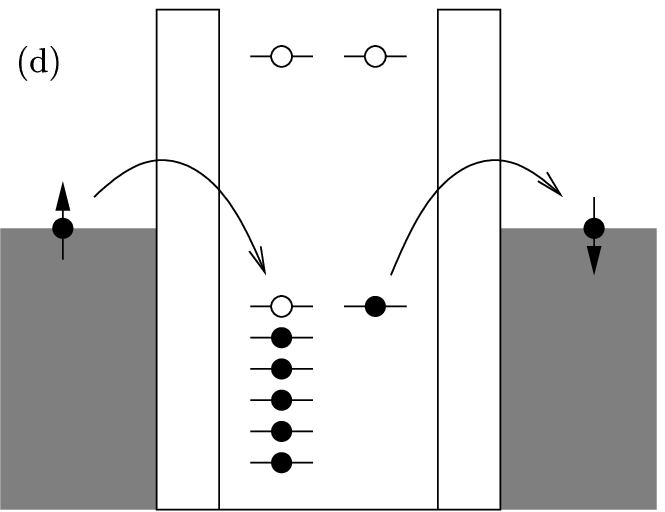}\\%
\includegraphics[width=30mm,clip=]{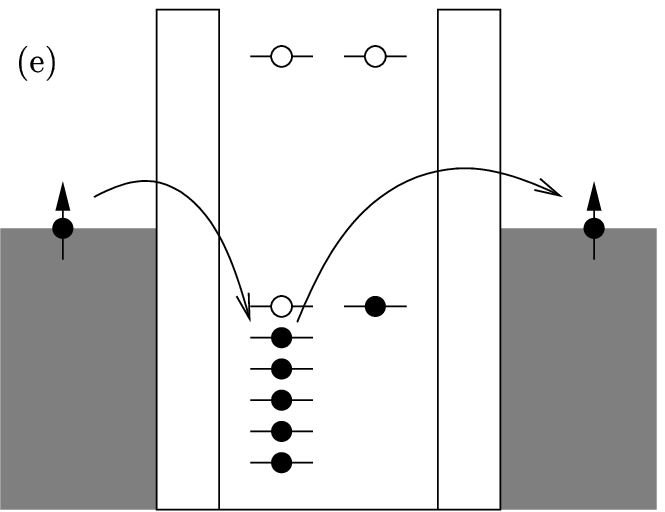}%
\caption{Virtual processes mediating the transitions (a) from $\ket{N,0}$
  to $\ket{N,0}$, (b) from $\ket{N,0}$ to $\ket{N-1,1}$, (c) from
  $\ket{N-1,1}$ to $\ket{N-1,1}$, (d) from $\ket{N-1,1}$ to
  $\ket{N,0}$, and (e) from $\ket{N-1,1}$ to $\ket{N-1,1}$.  The
  virtual process (e) does not have a counterpart in the usual
  Anderson model and gives an extra kinetic energy gain for the
  electrons in the state $\ket{N-1,1}$.}
\label{Kondo:fig2}
\end{figure}

Given the setup prescribed above, transport of the conduction
electrons through the quantum dot via sequential tunneling is highly
suppressed.  The hybridization of the dot levels are possible only
through the virtual tunneling processes.  To simplify the discussion,
we will further assume that $\mu_N-E_F\gg E_F-\mu_{N-1}$. The dominant
contributions then come from the virtual processes involving the
$(N-1)$-electron states $\ket{N-1,0}$ and $\ket{N-2,1}$, while the
processes involving the virtual states with $(N+1)$ electrons in the
dot can be ignored. [In the NRG study to be discussed below, however,
we have taken into account all the virtual processes with $(N-1)$- and
$(N+1)$-electron states.]  Effectively, the tunneling Hamiltonian
takes the form (see also Fig.~\ref{Kondo:fig1})
\begin{multline}
\label{Kondo:HT2}
H_T = \sum_kV\Bigl( c_{k\up}^\dag\ket{N-1,0}\bra{N,0} \\\mbox{} +
c_{k\down}^\dag\ket{N-1,0}\bra{N-1,1} \\\mbox{} -
c_{k\up}^\dag\ket{N-2,1}\bra{N-1,1} \Bigr) + h.c.
\end{multline}
Each term in Eq.~(\ref{Kondo:HT2}) leads to one of the virtual
processes listed in Fig.~\ref{Kondo:fig2}.  Within the spirit of the
Schrieffer-Wolf transformation taking into account these virtual
processes, one can show that for small $\Gamma$ and at low
temperatures, the impurity model $H=H_D+H_c+H_T$ is equivalent to a
Kondo-like model:
\begin{multline}
\label{Kondo:Kondo}
H = -\sum_{k,q} J_\bot\left( S^+\, c_{q,\up}c_{k,\down}^\dag + S^-\,
  c_{q,\down}c_{k,\up}^\dag \right) \\\mbox{} - \sum_{k,q}S^z\left(
  J_\up\, c_{q,\up}c_{k,\up}^\dag - J_\down\,
  c_{q,\down}c_{k,\down}^\dag \right) \,,
\end{multline}
where we have adopted the notation $S^+=\ket{N,0}\bra{N-1,1}$,
$S^-=\ket{N-1,1}\bra{N,0}$, and
$S^z=(\ket{N,0}\bra{N,0}-\ket{N-1,1}\bra{N-1,1})/2$ to emphasize the
roles of the states $\ket\Up\equiv\ket{N,0}$ and
$\ket\Down\equiv\ket{N-1,1}$ as pseudo-spin components.  The coupling
constants $J_\bot$, $J_\down$, and $J_\up$ in Eq.~(\ref{Kondo:Kondo})
are given by
\begin{equation}
J_\bot = \frac{|V|^2}{E_F+E_{N-1,0}-E_{N,0}-eV_g} \,,
\end{equation}
\begin{multline}
J_\up = \frac{|V|^2}{E_F+E_{N-1,0}-E_{N,0}-eV_g} \\ \mbox{} -
\frac{|V|^2}{E_F+E_{N-2,1}-E_{N-1,1}-eV_g} \,,
\end{multline}
and
\begin{equation}
J_\down = \frac{|V|^2}{E_F+E_{N-1,0}-E_{N-1,1}-eV_g} \,.
\end{equation}

Before going further, it will be useful to discuss here an important
difference between our impurity problem, the model studied in
Ref.~\onlinecite{Pustilnik00a}, and the usual Anderson model.
In the Anderson model, the degenerate states are spin states
associated with a single \emph{orbital} in the impurity.  The impurity
spin is isotropic. In particular, the spin components $\up$ and
$\down$ are equal in the contributions to the transport.
In our case, the pseudo-spin component $\ket\Down$ ($\ket{N-1,1}$)
allows for one additional virtual-process channel compared with
$\ket\Up$, namely, the one depicted in Fig.~\ref{Kondo:fig2} (e),
coming from the third term in the parenthesis in
Eq.~(\ref{Kondo:HT2}).  This additional channel is responsible for the
difference between the coefficients $J_\down$ and $J_\up$ in
Eq.~(\ref{Kondo:Kondo}).  Physically, this additional channel allows
the electrons in the $\ket\Down$ states for more kinetic energy gain
than those in the $\ket\Up$ states.  The difference
\begin{math}
\Delta_z \equiv J_\up - J_\down
\end{math}
leads to an \emph{effective} Zeeman splitting between $\ket\Up$ and
$\ket\Down$ states.  It is reminiscent of the original Kondo impurity
in an external magnetic field~\cite{Meir93a,Wingreen94a,corresp}.  It
should be emphasized, however, that in our case the fictitious field
$\Delta_z$ arises \emph{intrinsically}.  Moreover, as we see below,
the effects of $\Delta_z$ can be removed by detuning the levels of
$\ket{N,0}$ and $\ket{N-1,1}$ (i.e., going off the phase boundary in
the parameter space).
The two degenerate states involved in the Kondo effect studied in
Ref.~\onlinecite{Pustilnik00a} become degenerate due to competition
between the single-particle orbital-level spacing and the Zeeman
energy.
In the case of the Kondo effect near the singlet-triplet transition
studied in Ref.~\onlinecite{Eto00a,Izumida01a}, the degeneracy purely comes from
the many-body exchange interaction, and is four-fold (ignoring very
small Zeeman energy).  In our case, it is a result of interplay
between many-body interaction, parabolic confinement energy, and
(real) Zeeman effect.
We also remark that the tunneling amplitude $V$ in
Eq.~\eqref{Kondo/SchriefferWolf:HT} should, in general, depend on the
orbital $m$.  Its effect introduces another contribution to the
fictitious Zeeman splitting, and does not affect the results
qualitatively.

To investigate the low-energy physical properties of the Kondo-like
model, Eq.~(\ref{Kondo:Kondo}), we first follow the poor man's scaling
approach~\cite{Hewson93a} and trace out the conduction electrons in
the range $D-\delta{D}<|\veps_k|<D$.  The resultant renormalization
group (RG) equations are given by
\begin{equation}
\label{Kondo:RG}
\frac{d}{d\ell}J_\bot = 2\rho_0J_zJ_\bot \,,\quad
\frac{d}{d\ell}J_z = 2\rho_0J_\bot^2 \,,\quad
\frac{d}{d\ell}\Delta_z = 0 \,,
\end{equation}
where
\begin{math}
J_z \equiv (J_\down+J_\up)/2
\end{math}
and
\begin{math}
\ell \equiv -\ln(\rho_0D)
\end{math}.
The first two RG equations in Eq.~(\ref{Kondo:RG}) are exactly the
same as those of the usual Anderson model, which exhibit a
Kosterlitz-Thouless-type RG flow diagram~\cite{Hewson93a}.  This
implies that the effective impurity model in question exhibits a Kondo
effect.  The additional coupling constant $\Delta_z$ in
Eq.~(\ref{Kondo:RG}) plays a role of a fictitious magnetic field on
the pseudo-spin.
Therefore, for $\Delta_z$ larger than the Kondo temperature $T_K$, the
Kondo resonance peak at the Fermi level will be splitted into two
peaks with separation given by $\Delta_z$.  The spectral weights of
the splitted peaks will be diminished~\cite{Meir93a,Wingreen94a}.
Further, since $\Delta_z$ arises intrinsically in our model, the
effect of this fictitious field may be removed by detuning the levels
of $\ket{N,0}$ and $\ket{N-1,1}$.  All these arguments are confirmed
by the NRG calculations, see below.

\begin{figure}
\centering%
\includegraphics[width=55mm,clip=]{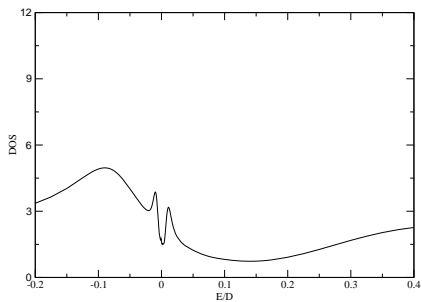}
\caption{Density of states on the quantum dot in a strong magnetic
  field with degenerate energy levels $E_{6,0}=E_{5,1}$.}
\label{Kondo:fig3}
\end{figure}

\begin{figure}
\centering%
\includegraphics[width=55mm,clip=]{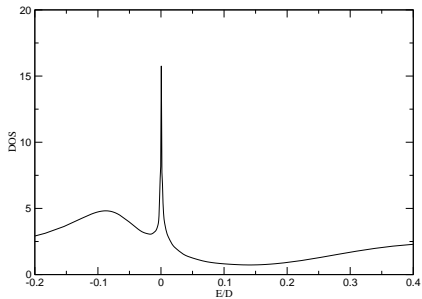}
\caption{Density of states on the quantum dot in a strong magnetic
  field with small detuning of energy levels $E_{6,0}$ and $E_{5,1}$.}
\label{Kondo:fig4}
\end{figure}

The RG equation, Eqs.~(\ref{Kondo:RG}), and the corresponding
arguments above are based on a second-order perturbation in $J_\bot$,
$J_\down$, and $J_\up$.  One may question the validity of the
perturbative RG analysis, in particular, the marginal behavior of
$\Delta_z$ in Eq.~(\ref{Kondo:RG}).  We perform a NRG
calculation~\cite{Bulla01a,Krishna-murthy80b,Krishna-murthy80a,Wilson75a},
and justify that the above conclusions are qualitatively correct.
Figure \ref{Kondo:fig3} shows the local density of states of the
quantum dot in a strong magnetic field.  For the calculations we have
chosen $\Gamma=0.025D$ and $E_F-(E_{N,0}-E_{N-1,0}+eV_g)=0.085D$.
Near the Fermi level, there are two Kondo peaks separated by
$\Delta_z\approx 0.01D$.  Since this fictitious Zeeman splitting comes
from the differences in the kinetic energy gains for $\ket\Up$ and
$\ket\Down$ states, it can be removed by detuning the degenerate
levels $E_{N,0}$ and $E_{N-1,1}$.  In Fig.~\ref{Kondo:fig4}, we have
detuned the energy levels of $E_{N,0}$ and $E_{N-1,1}$ by amount
$\Delta_z$, i.e.,
\begin{math}
E_F - (E_{N,0}-E_{N-1,0}+eV_g) = (0.085-0.005)D
\end{math}
and
\begin{math}
E_F - (E_{N-1,1}-E_{N-1,0}+eV_g) = (0.085+0.005)D
\end{math}.
One can clearly see that the Kondo resonance peak at the Fermi level
has been recovered.

Before concluding we briefly discuss possible experiments.  The
maximum density droplet states have already been probed in transport
measurements through vertical~\cite{Oosterkamp99a} and
lateral~\cite{Klein95a} quantum dots.  The ordinary Kondo effects have
also been observed in quantum
dots~\cite{Goldhaber-Gordon98a,Cronenwett98a,vanderWiel00a}.
To observe a Kondo effect involving the maximum density droplet
states, vertical quantum dots are particularly convenient.  One
advantage is that they provide a relatively strong confinement
potential.  Another important advantage of vertical dots is that the
leads can be made of bulk electrodes.  In bulk electrodes, the
densities of states for both spins near the Fermi level are
insensitive to the energy (we assume the Zeeman energy is small
compared with the Fermi energy) and the spin-polarization effects can
be safely neglected~\cite{vanderWiel00a}.
We estimate the typical values of the relevant parameters to be $N\sim
15$, $B\sim 5\T$ ($g\mu_BB\sim 0.15\meV$), $\hbar\Omega\sim 3\meV$,
$T_K\sim 0.05\mbox{--}0.1\K$, and $\Delta_z\sim 0.1\meV$.

We have reported on the Kondo effect of quantum dots in the quantum
Hall regime, involving the precursor of the Landau level filling
factor $\nu=1$ state.  Unlike the ordinary Anderson impurity model, we
find that the Kondo resonance splits due to an internally generated
fictitious magnetic field.  We showed that by detuning the energies of
states involved in the Kondo effect, the resonance peak at the Fermi
level may be restored.  We have argued that these effects may be
observed in vertical dots.

We thank Hangmo Yi for very useful discussions.  M.-S.C. acknowledges
the supports from the Swiss-Korean Outstanding Research Efforts Award
(SKORE-A) program and by the KOSEF through the eSSC at POSTECH.
S.R.E.Y was supported by the KOSEF Quantum-functional Semiconductor
Research Center at Dongguk University and by Grant No. R01-1999-00018
from the interdisciplinary Research program of the KOSEF.


\begin{thebibliography}{99}
\bibitem{corresp} There is one-to-one correspondence between our
  impurity problem and the common Anderson model (in the $U\to\infty$
  limit):
\begin{math}
\eps_d \leftrightarrow E_F - (E_{N,0}-E_{N-1,0}+eV_g)
\end{math},
\begin{math}
\ket{0} \leftrightarrow \ket{N-1,0}
\end{math},
\begin{math}
\ket\up \leftrightarrow \ket\Up
\end{math},
and
\begin{math}
\ket\down \leftrightarrow \ket\Down
\end{math}.
\end{thebibliography}

\end{document}